\documentclass[twocolumn]{aastex61}

\usepackage{graphicx}	
\usepackage{amsmath}	
\usepackage{amssymb}	
\usepackage{natbib}
\usepackage{xspace} 

\newcommand {\chandra} {{\it Chandra}\xspace}
\newcommand {\xmm} {{\it XMM-Newton}\xspace}

\newcommand {\swiftxrt} {{\it Swift}~XRT\xspace}
\newcommand {\swiftbat} {{\it Swift}~BAT\xspace}

\newcommand {\feka} {Fe~K$\alpha$\xspace}

\newcommand {\nh} {$N_{\mathrm{H}}$\xspace}

\newcommand {\Lbol} {$L_{\rm bol}$\xspace}

\newcommand{\oiii}{\mbox{[\ion{O}{3}]}\xspace}

\newcommand {\ergpersec} {erg~s$^{-1}$\xspace}

\newcommand {\nhunit} {cm$^{-2}$\xspace}
\newcommand {\degrees} {$^{\circ}$\xspace}

\newcommand {\xiunit} {erg~cm~s$^{-1}$\xspace}

\newcommand {\apec} {$\mathtt{apec}$\xspace}
\newcommand {\photemis} {$\mathtt{photemis}$\xspace}

\shorttitle{The Teacup in X-rays}
\shortauthors{Lansbury et al.}

\begin{document}

\title{Storm in a Teacup: X-ray view of an obscured quasar and superbubble}

\author[0000-0002-5328-9827]{George~B.~Lansbury}
\affil{Institute of Astronomy, University of Cambridge,
  Madingley Road, Cambridge, CB3 0HA, UK\ }
\altaffiliation{gbl23@ast.cam.ac.uk}

\author{Miranda~E.~Jarvis}
\affil{Max-Planck Institut für Astrophysik, Karl-Schwarzschild-Str. 1, 85741 Garching, Germany}
\affil{European Southern Observatory, Karl-Schwarzschild-Str.\ 2,
  85748 Garching, Germany}

\author{Chris~M.~Harrison}
\affil{European Southern Observatory, Karl-Schwarzschild-Str.\ 2,
  85748 Garching, Germany}

\author{David~M.~Alexander}
\affil{Centre for Extragalactic Astronomy, Department of Physics,
  Durham University, South Road,
  Durham, DH1 3LE, UK}

\author{Agnese~Del~Moro}
\affil{Max-Planck-Institut f\"ur Extraterrestrische Physik
  (MPE), Postfach 1312, 85741, Garching, Germany}

\author{Alastair~C.~Edge}
\affil{Centre for Extragalactic Astronomy, Department of Physics,
  Durham University, South Road,
  Durham, DH1 3LE, UK}

\author{James~R.~Mullaney}
\affil{Department of Physics and Astronomy, The University of
  Sheffield, Hounsfield Road, Sheffield, S3 7RH, UK}

\author{Alasdair~P.~Thomson}
\affil{Jodrell Bank Centre for Astrophysics, School of Physics and
  Astronomy, The University of Manchester, Alan Turing Building,
  Upper Brook Street, Manchester M13 9PL, UK}



\begin{abstract}
We present the X-ray properties of the ``Teacup AGN'' (SDSS\,J1430+1339), a
$z=0.085$ type~2 quasar which is 
interacting dramatically with its host galaxy. Spectral modelling
of the central quasar reveals a powerful, highly obscured AGN with a
column density of $N_{\rm H}=(4.2$--$6.5)\times 10^{23}$~\nhunit and an intrinsic 
luminosity of $L_{\rm  2\mbox{-}10\,keV}=(0.8$--$1.4)\times
10^{44}$~\ergpersec.  The current high bolometric luminosity inferred
($L_{\rm bol}\approx 10^{45}$--$10^{46}$~\ergpersec) has ramifications for
previous interpretations of the Teacup as a fading/dying quasar. 
High resolution \chandra imaging data reveal a $\approx 10$~kpc loop of X-ray emission, 
co-spatial with the ``eastern bubble'' previously identified in
luminous radio and ionised gas (e.g., \oiii line) emission. 
The X-ray emission from this structure is in good agreement with a shocked thermal gas, with
$T=(4$--$8)\times 10^{6}$~K, and there is evidence for an
additional hot component with $T\gtrsim 3\times 10^{7}$~K.
Although the Teacup is a radiatively dominated AGN, the estimated
ratio between the bubble power and the X-ray luminosity is in
remarkable agreement with observations of ellipticals, 
  groups, and clusters of galaxies undergoing AGN feedback.
\end{abstract}




\section{Introduction}

There is now general agreement that growing supermassive black holes,
i.e., active galactic nuclei (AGN), have a dramatic impact upon the
evolution of their host galaxies. Observationally constraining the
details of this so-called ``AGN feedback'' is an ongoing challenge of
extragalactic astronomy (e.g.,
\citealt{Fabian12,Heckman14,Harrison17}). There is convincing evidence
that powerful radio jets launched by some AGN, whose energetic output
is dominated by mechanical energy, are effective at regulating the
cold gas supply to their host galaxies
(e.g., \citealt{McNamara12}). However, it is still unclear how radiatively
dominated (or ``radio quiet'') AGN interact with their host
galaxies (e.g., \citealt{Mullaney13,Greene14,Zakamska14}), despite these being
the majority population (e.g., $\sim 90\%$ of quasars; \citealt{Zakamska04}).

One low-redshift Type~2 quasar ($z=0.0852$) that has received
attention in the literature is SDSS~J143029.88 +133912.0. The object
has been labelled the ``Teacup AGN''
due to a ``handle'' of ionized gas extending $\approx 10$~kpc to the
east of the galaxy core, initially identified in SDSS images by Galaxy Zoo
(Massimo Mezzoprete, 2007; \citealt{Lintott08}) and later confirmed by
{\em HST} imaging (\citealt{Keel12a,Keel15}). Karl G.~Jansky Very Large Array (VLA)
data subsequently revealed that radio emission traces this ``eastern bubble'', and
that this is just one of two bi-polar radio superbubbles, extending to
distances of $\approx 10$~kpc in opposite directions
(\citealt{Harrison15}). A $\approx1$\,kpc small jet-like feature was
also identified, spatially co-incident with a $\approx700$\,km\,s$^{-1}$
outflow of warm ionised gas, along the same orientation axis as the eastern bubble (\citealt{Harrison15}).

On the one hand, the Teacup has been used as a case study for how typical
quasars may interact with their host galaxies
(\citealt{Harrison14,VillarMartin14,Harrison15,RamosAlmeida17}). On
the other hand the Teacup has been identified as a ``fading''/``dying'' quasar,
like Hanny’s Voorwerp (e.g., \citealt{Keel12b,Sartori18}),
where the ionisation state of the extended optical emission line region
provides evidence that the AGN was more luminous in the past
(\citealt{Gagne14,VillarMartin18,Keel17}).  In both cases it is
crucial to have an excellent constraint on the current bolometric
luminosity of the central engine - in order to assess (a) the
available energy for feedback and (b) the magnitude of AGN fading
over time. Furthermore, understanding the physical processes in the
bubble is essential for testing both the feedback and fading scenarios. Towards
addressing these issues, here we present X-ray constraints
on both the central source and the extended bubble of the Teacup.
We use ($\Omega_{M}$, $\Omega_{\rm \Lambda}$,
$h$)$=$($0.27$, $0.73$, $0.70$).

\section{Observations}
\label{sec:data}

The X-ray observations of the Teacup 
are summarized in Table \ref{tab:xrayData}.
The first X-ray detection of the Teacup was obtained with the XRT
instrument on the {\it Neil Gehrels \it Swift} observatory, in
directors' discretionary time observations (PI: Del~Moro). Only $\approx
19$ photons were detected, but the data nevertheless hinted at a
highly obscured central quasar. 
A deeper followup observation was then performed with \chandra ($48.4$~ks; PI:
Harrison), using the Advanced CCD Imaging Spectrometer (ACIS). 
We process the data using the CIAO tool
$\mathtt{chandra\_repro}$.\footnote{http://cxc.harvard.edu/ciao/ahelp/chandra\_repro.html} 
As there are no periods of significant background flaring, we use the
full exposure. The \chandra data are the main constituent of this work,
used to analyse both the spatially resolved X-ray emission and the central quasar.
For the spectral analysis of the quasar, we additionally use data from an
\xmm observation of the Teacup (PI: Maksym), processed using the \xmm
Science Analysis Software (SAS v.15.0.0).

\renewcommand*{\arraystretch}{1.7}
\begin{table}
\centering
\caption{X-ray data for the Teacup}
\begin{tabular}{llllll} 
\hline\hline \noalign{\smallskip}
Observatory & Obs ID & UT Date & $t$ & $S_{\rm net}$ & $B$ \\
(1) & (2) & (3) & (4)  & (5)  & (6) \\
\noalign{\smallskip} \hline \noalign{\smallskip}
\parbox[t]{1.7cm}{{\it Chandra}\,ACIS} & \parbox[t]{1.4cm}{18149} & \parbox[t]{1.6cm}{2016-04-19} & \parbox[t]{0.4cm}{48.4} & \parbox[t]{0.4cm}{702} & \parbox[t]{0.4cm}{0} \\
\parbox[t]{1.7cm}{{\it XMM}\,PN {\it XMM}\,MOS} & \parbox[t]{1.4cm}{0762630101 $\cdots$} & \parbox[t]{1.6cm}{2016-02-10 $\cdots$} & \parbox[t]{0.4cm}{26.7 55.8} & \parbox[t]{0.4cm}{1625 1862} & \parbox[t]{0.4cm}{157 125} \\
\parbox[t]{1.7cm}{{\it Swift}\,XRT} & \parbox[t]{1.4cm}{00033655$^{\mathrm{a}}$} & \parbox[t]{1.6cm}{2015-03} & \parbox[t]{0.4cm}{5.5} & \parbox[t]{0.4cm}{19} & \parbox[t]{0.4cm}{1} \\
\noalign{\smallskip} \hline \noalign{\smallskip}
\end{tabular}
\begin{minipage}[c]{0.97\columnwidth}
\footnotesize
\textbf{Notes.} (1): Observatory and instrument. (2) and
(3): Observation ID and start date,
respectively. (4): Net exposure time (ks). (5): Total net source
counts in the full-band (i.e., $0.5$--$8$, $0.5$--$10$,
 and $0.6$--$10$~keV for \chandra, \xmm, and \swiftxrt). 
(6): Background counts in the source region. 
\end{minipage}
\label{tab:xrayData}
\end{table}

\section{Results}

\subsection{X-ray imaging}
\label{sec:xray_imaging}

\begin{figure*}
	\includegraphics[width=\textwidth]{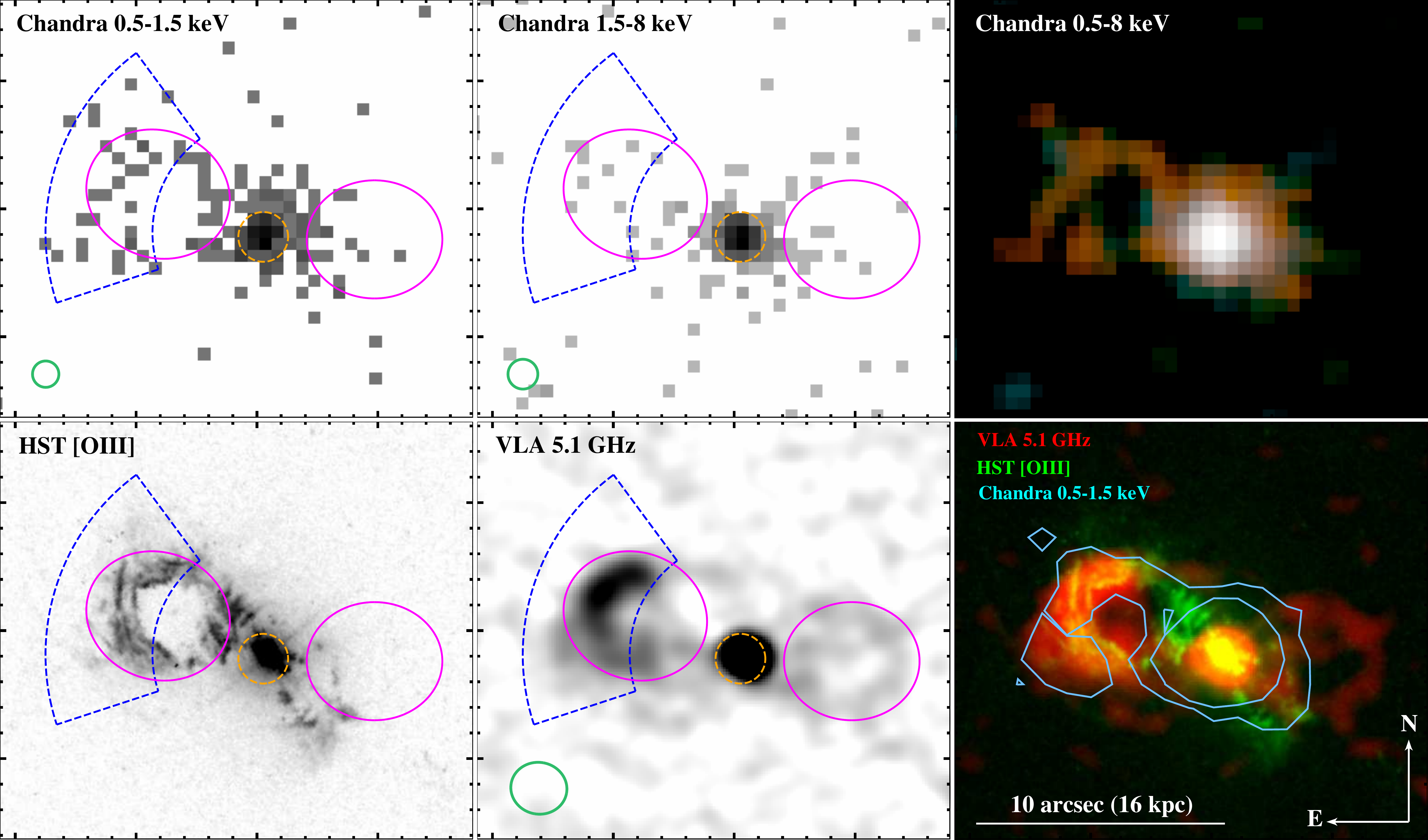}
    \caption{Images of the Teacup from
      \chandra (top row), the \oiii narrow band with {\it HST} (lower
      left; \citealt{Keel15}), and the VLA at $5.12$\,GHz (lower
      middle; \citealt{Harrison15}). 
      Magenta elliptical regions highlight the bipolar radio
      bubble structures. The spectral extraction regions (Section
      \ref{sec:xray_spectral}) are shown in orange and blue. 
      Green ellipses show PSF half-energy widths. 
      Upper right: \chandra color composite (smoothed with a 1$''$
      gaussian) where red, green, and blue correspond to $0.5$--$1.5$,
      $0.5$--$8$, and $1.5$--$8$\,keV, respectively. 
      Lower right: color composite of the radio (red) and
  \oiii (green) images, with cyan contours showing the spatial
  distribution of $0.5$--$1.5$\,keV emission.}
    \label{fig:images}
\end{figure*}

Here we use the \chandra imaging (Figure \ref{fig:images})
to identify spatially distinct components of the X-ray emission. 

The brightest component is the central quasar, which lies at a 
position of $\mathrm{14\mbox{:}30\mbox{:}29.897}$
$\mathrm{+13\mbox{:}39\mbox{:}11.88}$ in the \chandra full-band image.
  Close to the quasar we find 
  evidence, from ray-tracing PSF simulations (with $\mathtt{MARX}$;
  \citealt{Davis12}), for spatially extended $E\lesssim 1.5$\,keV emission
  on scales of $\approx 0.6$--$2.5''$ ($1$--$4$~kpc). This central extended emission is not a main focus of
  this paper, but it motivates the relatively small source
  extraction region adopted in Section \ref{sec:xray_spec_agn}.

To the east of the central quasar, there is a
striking loop of low-energy ($\approx 0.5$--$1.5$\,keV) emission extending out to a maximum distance
of $\approx 7.5''$ ($12$~kpc). This X-ray emission traces the
morphology of luminous radio emission and ionized gas (e.g., \oiii) in
the ``eastern bubble'' of the Teacup. 
At higher energies, the bubble is faint but still significant; the photon 
counts at $1.5$--$8$ and $2$--$8$\,keV suggest 
binomial false probabilities (e.g., \citealt{Weisskopf07}) of
$P_{\rm false}=8\times 10^{-5}$ and $0.005$, respectively. 
We also note that there is tentatively some extended
  high-energy ($1.5$--$8$\,keV) emission around the {\it western}
  radio bubble, but this has a low significance ($P_{\rm false}=0.1$).

\subsection{X-ray spectral properties}
\label{sec:xray_spectral}

\subsubsection{The central quasar}
\label{sec:xray_spec_agn}

\begin{figure*}
 \center
	\includegraphics[width=0.9\textwidth]{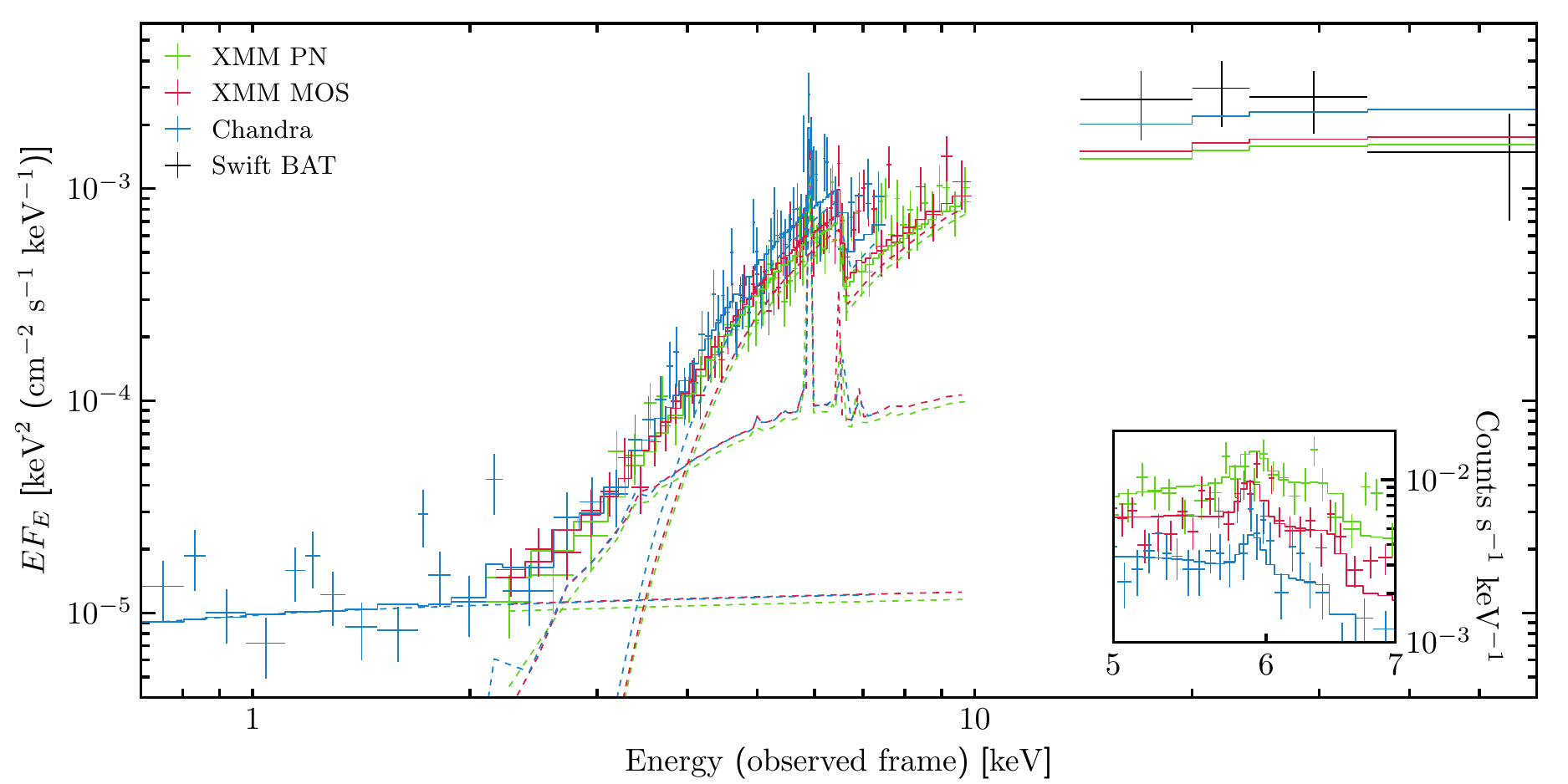}
    \caption{X-ray spectra for the central quasar. 
      The solid curves show the $\mathtt{borus02}$ model best
        fit to the \chandra and \xmm spectra, and the dashed lines show the model subcomponents. 
      \swiftbat data, not used in the fitting, are 
        shown for comparison. Inset: zoom-in of the \feka line, in
      count-rate units. 
}
    \label{fig:xray_spec_agn}
\end{figure*}

For the quasar, we extract source spectra from the \chandra and
\xmm (PN and MOS) data sets, using circular regions of $1''$ and $15''$ radius,
respectively. 
For \xmm, we 
limit the analysis to $E>2$\,keV, where the central point source
dominates, based on the high-resolution \chandra data. 
The background spectra are extracted from large source-free regions on
the same detector chip as the source.
  The background-subtracted source spectra are shown
  in Figure \ref{fig:xray_spec_agn}. 
  At $2$--$8$\,keV, the \chandra flux is a factor of
  $1.31^{+0.11}_{-0.14}$ higher than {\it XMM}, 
  in agreement with typical AGN X-ray variability on multi-month
  timescales (e.g., \citealt{Yang16,Lansbury17a,Ricci17d}).
  We model the spectra simultaneously using {\sc XSPEC} (version 12.9.1p;
  \citealt{Arnaud96}), adopting $\chi^{2}$ statistics for fitting, and
  grouping by a minimum of 10 and 20 counts per spectral bin for \chandra
  and \xmm, respectively. 
  
  The X-ray spectrum rises to higher energies with an observed photon
  index of $\Gamma_{\rm 3\mbox{-}8\,keV}\approx -0.74$, 
  suggesting obscuration by a high column density (\nh) of gas.
   We thus fit the spectrum with an absorbed power law
  model ($\mathtt{cabs}\cdot\mathtt{zwabs}\cdot\mathtt{pow}$ in {\sc
    XSPEC} formalism). 
  For the intrinsic power law ($\mathtt{pow}$), we fix the photon
  index to a representative value of $\Gamma=1.9$ and we
  allow different normalizations for \chandra and \xmm, given the
  factor of $1.31$ mentioned above.
Two additional model components are required: 
  an unobscured power law to parameterize the
  low-energy ($\lesssim 2$\,keV) emission, consistent
  with electron-scattering of the primary AGN power-law; 
  and a narrow redshifted gaussian to fit \feka line emission 
    at rest-frame $E=6.4$\,keV. This line component is required at
    the $4.1\sigma$ significance level, based on 50,000 simulations of
    the continuum-only spectrum (using $\mathtt{simftest}$ in {\sc XSPEC}).
    We find no statistical requirement for adding ionized \feka line
    components to the model (e.g., at $6.7$\,keV).

  The best fit has $\chi^{2}/n=1.18$, and 
  the column density is well constrained as $N_{\rm H}=(4.5\pm
  0.3)\times 10^{23}$~\nhunit. The \feka equivalent width of
    $\mathrm{EW_{FeK\alpha}}=0.12\pm 0.03$~keV, measured over the
    absorbed power-law continuum at $E>4$\,keV, is consistent with
    expectations from the sub-Compton-thick column density.
  Correcting for absorption, the AGN is intrinsically luminous, with a rest-frame
  $2$--$10$\,keV luminosity of $L_{\rm 2\mbox{-}10}=(0.8$--$1.0)\times 10^{44}$~\ergpersec. 

 The above \nh is sufficiently high that Compton-scattered (i.e., ``reflected'')
  continuum emission may contribute to the X-ray spectrum. We therefore
  additionally fit a physically motivated torus model ($\mathtt{borus02}$; 
  \citealt{Balokovic18}), to self-consistently model primary and 
  reflected emission. A statistically acceptable fit is obtained in a
  basic model setup where the line-of-sight column density (\nh) is untied
  from the average toroidal column density ($N_{\rm H,tor}$), as expected for a
  non-uniform torus, and standard values are assumed for the iron
  abundance (set to solar), the torus covering factor (fixed at
  $0.5$), and the inclination angle (fixed at $87$\degrees).
  The best fit has $\chi^{2}/n=1.12$, $N_{\rm
    H}=(5.9^{+0.6}_{-0.5})\times 10^{23}$~\nhunit, $N_{\rm
    H,tor}=(1.1^{+0.2}_{-0.3})\times 10^{23}$~\nhunit,
  and $L_{\rm 2\mbox{-}10}=(1.1$--$1.4)\times 10^{44}$~\ergpersec. 
  
  Adding further confidence to these \nh and $L_{\rm 2\mbox{-}10}$
  values, the Teacup is detected in the \swiftbat $105$-month survey (\citealt{Oh18})
  and the BAT spectrum and luminosity ($L_{\rm
    14\mbox{-}195}=1.8\times 10^{44}$~\ergpersec) are consistent
  with the above best-fit models (Figure \ref{fig:xray_spec_agn}). The
  agreement with the BAT spectrum
  (obtained over $\approx 9$~years, up to 2013) 
  suggests that there has been no overall trend of brightening or
  fading during the full $\approx 11$-year
  period of X-ray coverage, and that the Teacup has maintained a
  relatively constant average luminosity (within a factor of $\lesssim 2$).

\subsubsection{The eastern bubble}
\label{sec:xray_spec_bubble}

\begin{figure}
	\includegraphics[width=\columnwidth]{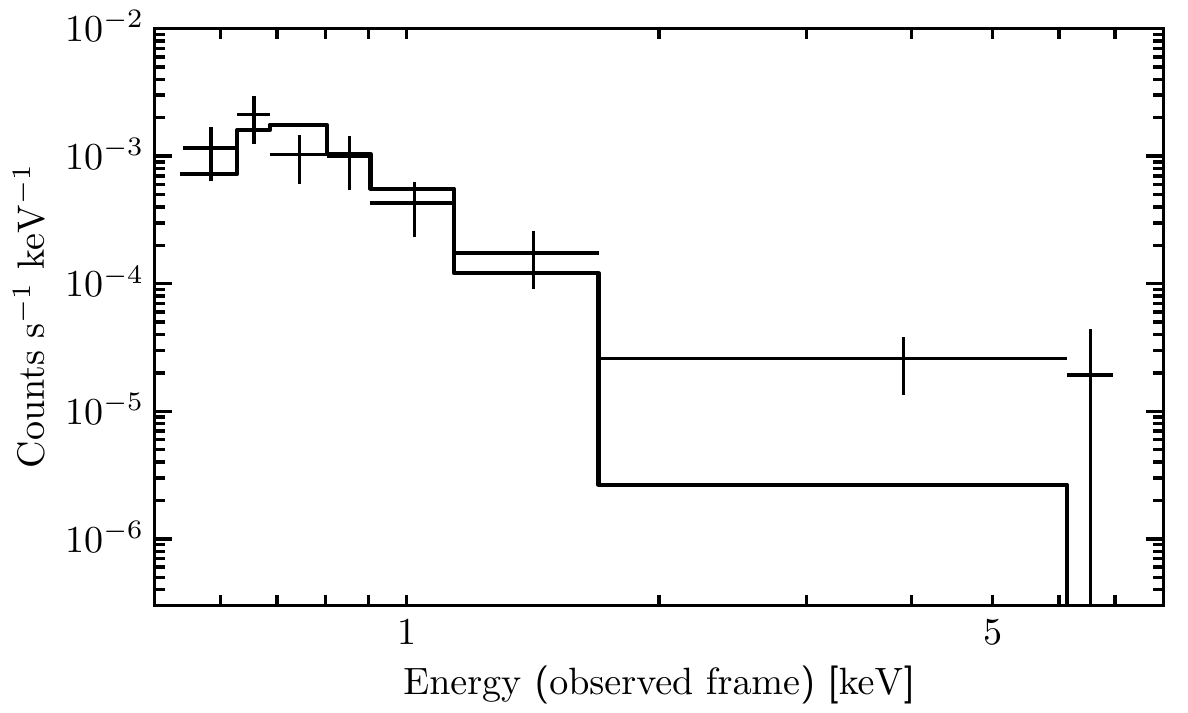}
    \caption{\chandra spectrum for the eastern bubble, and the best-fit
      \apec model. Each bin has a minimum significance of
      $2\sigma$, for visual purposes.}
    \label{fig:xray_spec_bubble}
\end{figure}

  For the eastern bubble, 
  we extract a \chandra spectrum from the region shown in
  Figure \ref{fig:images}, an annular sector at $4.7$--$9''$ from
  the central quasar. The region has minimal contamination
  from the quasar PSF ($\approx 0.5$ photons, based on
  $\mathtt{MARX}$ simulations).
The photon counts
are low, with $37.5$ net source counts ($2.5$ background counts) at
$0.5$--$8$\,keV. We therefore adopt the $C$ statistic for
fitting,\footnote{https://heasarc.gsfc.nasa.gov/docs/xanadu/xspec/wstat.ps}
and group the spectrum by a minimum of $1$ count per bin.

  The bubble spectrum is very steep ($\Gamma_{\rm
    0.5\mbox{-}8}=4.9^{+1.1}_{-1.0}$; Figure
  \ref{fig:xray_spec_bubble}) and comparatively
  luminous at low energies, with $L_{\rm
    0.5\mbox{-}2}=(1.3^{+0.3}_{-0.5})\times 10^{41}$~\ergpersec, in
  agreement with
  thermal gas emission. We therefore apply the
  $\mathtt{apec}$ model for a collisionally ionized diffuse gas
  (\citealt{Smith01}), which yields a statistically acceptable fit to
  the data ($C/n=32/31$; Figure \ref{fig:xray_spec_bubble}), 
  and a normalization of $\eta=3.9\times
  10^{-6}$. 
  The gas temperature is constrained to be
  $kT=0.4^{+0.3}_{-0.1}$\,keV, or $T=(3.5$--$8.1)\times
  10^{6}$\,K. 
This simple single-phase model leaves some positive flux residuals
at $\gtrsim 2$\,keV, where the X-ray emission is only weakly detected
($\approx 99.5\%$ confidence level; Section \ref{sec:xray_imaging}).  
This excess could plausibly represent an additional
hotter gas phase, with $L_{2\mbox{-}10}\approx 1.5\times
10^{40}$~\ergpersec and $kT\gtrsim 2.7$\,keV (i.e., $T\gtrsim 3\times
10^{7}$\,K). 
  The excess is unlikely due to star formation, as the luminosity expectation for such a small
  volume of the galaxy is significantly lower than
  $10^{40}$~\ergpersec (e.g., \citealt{Aird17a}). 

 We also consider the possibility of a photoionized, rather than
 thermal, emitter. Based on fits with the $\mathtt{photemis}$ model 
 (\citealt{Kallman01}; which uses {\sc XSTAR}\footnote{https://heasarc.gsfc.nasa.gov/docs/software/xstar/xstar.html}
 tables), a photoionized gas is consistent with the bubble spectrum
 for an ionization parameter of $\log \xi=1.7^{+0.3}_{-0.4}$
 ($C/n=32/31$).

\section{Discussion}

\subsection{A highly obscured, luminous quasar}
\label{sec:luminosity}

We have shown that the Teacup quasar is {\it currently}
accreting at a high luminosity, especially once the line-of-sight
obscuration ($N_{\rm H}\approx 5\times 10^{23}$~\nhunit) is corrected
for. The measured X-ray luminosity of $L_{2\mbox{-}10}=
(0.8$--$1.4)\times 10^{44}$~\ergpersec implies a high bolometric
luminosity ($L_{\rm bol}$), which is crucial to consider when connecting the energetics of
the central engine to phenomena observed in the host galaxy. 

In particular, the Teacup has previously been interpreted as a fading/dying
quasar due to the observed high-ionization emission line
regions (e.g., around the eastern bubble). Photoionization modelling
of these regions suggests that the central AGN was more luminous
in the past. For instance, \citet{Gagne14} find a highest past
luminosity of $L_{\rm  bol}^{\rm past}\approx 2\times 10^{46}$~\ergpersec,
in agreement with the results of \citet{VillarMartin18} for
gas at larger distances ($>15$\,kpc).
The extent of any fading since then is crucially dependent on
knowledge of the {\it current} bolometric luminosity.
 Fading by factors of $\approx 50$--$600$ have been
 inferred to have occured over $\approx (0.4$--$1)\times 10^{5}$~yr timescales
  (e.g., \citealt{Gagne14,Keel17,VillarMartin18}), but this was 
  based on an apparently low current luminosity
  ($L_{\rm bol}\approx 2\times 10^{44}$~\ergpersec).

The X-ray bolometric correction ($\kappa_{\rm bol}=L_{\rm
  bol}/L_{\rm 2\mbox{-}10}$) is likely to lie in the range $10\lesssim 
\kappa_{\rm bol}\lesssim 100$, based on typical values for
AGN (e.g., \citealt{Vasudevan09a, Vasudevan09b,
  Lusso12}). We therefore find, based on our $L_{\rm 2\mbox{-}10}$ constraints, that
$8\times 10^{44}\lesssim L_{\rm bol}\lesssim 1.4\times
10^{46}$~\ergpersec.
This agrees with the luminosity inferred from mid--far-IR SED fitting, $L_{\rm bol}\approx 2\times
10^{45}$~\ergpersec (\citealt{Harrison14}).
Based on the possible \Lbol values, 
there is no strong requirement for the Teacup to have undergone 
a dramatic fading 
over $\approx 100\,000$~yr. 
Assuming an upper limit to the past
luminosity of $L_{\rm bol}^{\rm past}\lesssim 2\times
10^{46}$~\ergpersec, the required fading factor is $\lesssim
25$.

\subsection{Understanding the superbubble}

  The eastern bubble is luminous
  in radio emission and spatially coincident high-ionization
  optical-line emission. Here we have shown that luminous X-ray emission
  traces the same structure. 
  Below we discuss possible emission mechanisms to explain the
  X-ray properties.

\subsubsection{Photoionized gas}
\label{sec:discussion_photo}

As shown in Section \ref{sec:xray_spec_bubble}, a
photoionized gas model can describe the bubble X-ray spectrum, for an ionization
parameter of $\xi\approx 50$~\xiunit.  
The large distance ($R\approx 10$\,kpc) is problematic in terms
of ionization by the central quasar, however, as the gas density must be low, with
$n=L_{\rm X}/\xi R^{2}\approx 0.002\,\mathrm{cm^{-3}}$ (e.g., \citealt{Kallman82}).  
The implied volume of the gas, estimated from the \photemis emission measure ($\mathrm{EM}=\int
n_{\rm e}n_{\rm H}dV$), is then $\gtrsim$three orders of magnitude too large
compared to the observed volume of the bubble shell. 
If we assume the quasar had a higher past $L_{\rm X}$ of $10^{45}$~\ergpersec (i.e., the same as PDS\,456, the
most luminous AGN at $z<0.3$), the required volume is still more than an
order of magnitude too high. We therefore prefer the solution
presented in Section \ref{sec:discussion_thermal}.

\subsubsection{Thermal gas resulting from an AGN outflow}
\label{sec:discussion_thermal}

The Teacup radio bubbles were likely formed by 
  jets or quasar winds from the central AGN (e.g.,
  \citealt{Harrison15}), which we have confirmed to be intrinsically
  energetic (Section \ref{sec:luminosity}).
The coupling of such outflows with the interstellar medium can explain the
eastern bubble X-ray spectrum, which is consistent with a
collisionally dominated thermal gas with
$T=(3$--$8)\times 10^{6}$\,K (Section \ref{sec:xray_spec_bubble}). 
In the nearby Universe, there are lower luminosity
Seyferts with AGN-driven radio bubbles/lobes, where X-ray-emitting gas
is found with similar temperatures (e.g.,
\citealt{Croston08,Mingo11,Mingo12,Paggi12}). In
these cases the gas is thought to be shock-heated by AGN-driven
outflows. 
The Teacup also bears resemblance to the mechanically driven
bubbles/cavities observed at the centers of galaxy
groups and clusters (e.g., \citealt{Birzan04,Russell13}). These systems can be
morphologically similar to the Teacup in the radio (e.g., \citealt{Simionescu18})
and X-ray (e.g., \citealt{Fabian06}) bands individually, and there are
also a few examples where bright optical line-emitting gas encases the
bubble (e.g., \citealt{Canning13}). It is not common, however, to
observe such a tight spatial correlation between the bright radio and
X-ray emission (Figure \ref{fig:images}). 

We also find a hard X-ray excess in the bubble (Section \ref{sec:xray_spec_bubble}),
which may represent a fainter, hotter X-ray-emitting
gas phase, with $T\gtrsim3\times 10^{7}$\,K. 
This agrees with expectations for a tenuous but energetic 
quasar wind (e.g., \citealt{Strickland00,Greene14,Nims15,Costa18}) that itself shocks cooler gas clouds (e.g., at the edge
of the expanding bubble). The latter then emit strongly in low-energy ($T\approx
\mathrm{few}\times 10^{6}$\,K) X-rays and optical-line emission (e.g.,
\citealt{Strickland00}). However, deeper X-ray
  observations are required to confirm the nature and spatial distribution of this hard X-ray
  excess.

\begin{figure}
	\includegraphics[width=\columnwidth]{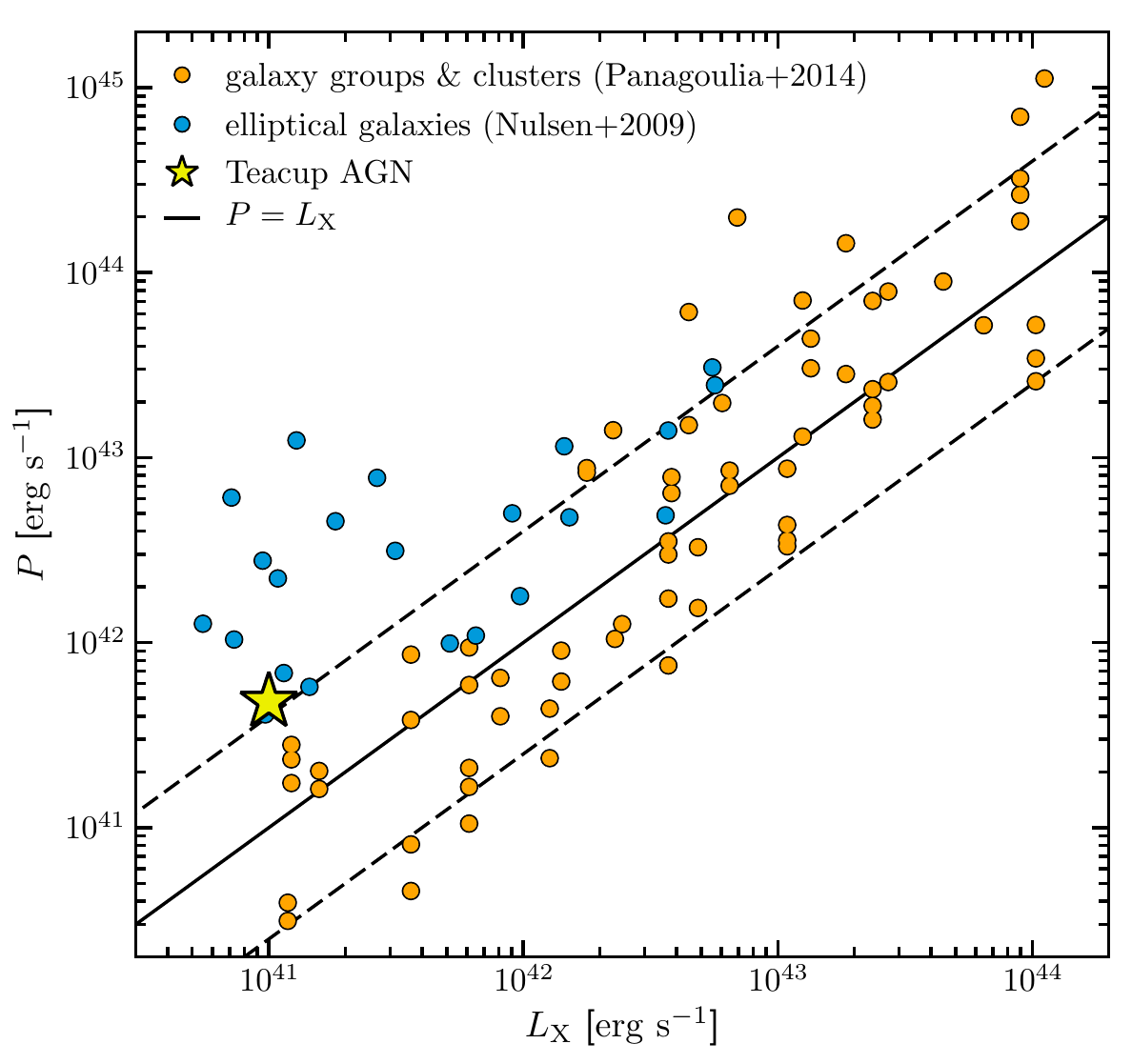}
    \caption{Bubble/cavity power against cooling X-ray luminosity. The
      Teacup bubble (yellow star) is
      compared with those in ellipticals (\citealt{Nulsen09}), galaxy groups, and clusters (\citealt{Panagoulia14}).}
    \label{fig:P_L}
\end{figure}

Here we consider the physical properties implied for the eastern bubble.
  Motivated by the X-ray, optical, and radio images, we assume that
  the gas occupies a $1.7$\,kpc thick spherical shell. The volume within the
  bubble extraction region (Figure \ref{fig:images}) is then
  $V=3\times 10^{66}\,\mathrm{cm}^{3}$.
  Given the \apec normalisation
  ($\eta$),\footnote{$\eta=10^{-14}(4\pi)^{-1}[D_{\rm
      A}(1+z)]^{-2}\int n_{\rm e}n_{\rm H}dV$, where $D_{\rm A}$ is
    the angular diameter distance.} this implies an electron
  density of $n_{\rm e}\approx 0.056$\,$\mathrm{cm^{-3}}$, and a
  gas mass of $M\approx 1.6\times 10^{8}\,M_{\odot}$.
  Taking the temperature of $kT=0.4$\,keV, and a particle number
  density of $n=1.92 n_{\rm e}$, the total thermal energy of the
  gas is large with $U=(3/2)nkTV\approx 3.1 \times 10^{56}$\,$\mathrm{erg}$. 
  Assuming that the energy injection from the central quasar has an
  age of $t=20$~Myr, based on the expected local sound
  speed,\footnote{The local speed of sound is determined as $c_{\rm
      s}=\sqrt{\gamma k T / \mu m_{\rm H}}$, where $\gamma=5/3$,
    $\mu=0.61$, and $m_{\rm H}$ is the atomic mass of hydrogen.} 
  the power of the bubble is $P\approx 4.9\times
  10^{41}$\,$\mathrm{erg\,s^{-1}}$. 
  Figure \ref{fig:P_L} compares this inferred power to the 
  X-ray luminosity of the bubble ($L_{\rm X}\approx
  10^{41}$~\ergpersec; Section \ref{sec:xray_spec_bubble}).
  As shown, the ratio $P/L_{\rm X}$ is in agreement with the 
  relationship between heating power and cooling X-ray luminosity found for
  bubbles/cavities in ellipticals, groups and clusters undergoing
  mechanically dominated AGN feedback (e.g.,
  \citealt{Fabian12,McNamara12,Panagoulia14}).
Although these bubbles do not typically resemble the Teacup
morphology across all wavelengths, Figure 4 
provides some evidence for similarities between the physical mechanisms of AGN
feedback in these systems.

\section{Summary}

In this work we have characterized the Teacup AGN and host galaxy, in
terms of their X-ray properties. The main results are as follows:

\begin{itemize}

\item The high resolution \chandra images reveal a striking loop of 
  X-ray emission, extending to $\approx 10$\,kpc. This is spatially
  coincident with the well-known eastern bubble, luminous in
  radio emission and ionized gas. 

\item Modelling the \chandra and \xmm spectra of the central
  quasar, we find it to be highly obscured, with $N_{\rm H}=(4.2$--$6.5)\times
  10^{23}$~\nhunit, and intrinsically luminous, with $L_{\rm  2\mbox{-}10\,keV}=
  (0.8$--$1.4)\times 10^{44}$~\ergpersec. 
  The quasar is inferred to be currently accreting at a high bolometric
  luminosity ($L_{\rm  bol}\approx 10^{45}$--$10^{46}$~\ergpersec).
  We therefore find no strong requirement for a fading/dying quasar
  scenario to explain the ionized gas emission in the eastern bubble. 

\item X-ray emission from the eastern bubble is spectrally consistent
  with a thermal (e.g., shock-heated) gas of temperature
  $kT=0.4^{+0.3}_{-0.1}$\,keV.  
  There is also evidence for a fainter component of
  $kT\gtrsim2.7$\,keV gas, in agreement with a quasar wind, but deeper observations are required
  to study this hotter gas phase. 

\item The ratio between the inferred bubble power ($P\approx
  4.9\times 10^{41}$~\ergpersec) and the X-ray luminosity ($L_{\rm
    X}\approx 10^{41}$~\ergpersec) is remarkably similar to that seen 
  for bubbles/cavities in ellipticals, galaxy
  groups, and clusters undergoing mechanically dominated AGN feedback.
  Our observations of the Teacup provide evidence of similar
  feedback processes at work in a quasar.

\end{itemize}

\section*{Acknowledgements}

This work was supported by a Herchel Smith Postdoctoral Research
Fellowship of the University of Cambridge (G.B.L.). 
Additional support came from the Science and Technology Facilities
Council (STFC) grant ST/L00075X/1 (D.M.A.\ and A.C.E.).
We thank the referee for the positive and constructive review.
We extend gratitude to Helen Russell, Dom Walton, Andrew Fabian, Peter
Kosec, and Rebecca Canning
for the useful discussions, and Julie Hlavacek-Larrondo and Electra
Panagoulia for providing data points from the literature.




\bibliographystyle{mnras}


\end{document}